\documentclass[12pt]{article}
\newcommand\be{\begin{equation}}
\newcommand\ee{\end{equation}}
\newcommand\ba{\begin{eqnarray}}
\newcommand\ea{\end{eqnarray}}
\newcommand\eq{\begin{equation}}           
\newcommand\en{\end{equation}}

\input epsf
\usepackage{amssymb,amsmath,graphicx} 
 
\begin{document}

\title{
{\hfill \small UMN-TH-2621/07 \\
\hfill \small FTPI-MINN-07/30
\\ ~\\~\\}
Sterile neutrino dark matter in warped extra dimensions
}
\author{Kenji Kadota
\\
\\
{\em  \small William I. Fine Theoretical Physics Institute, University of Minnesota, Minneapolis, MN 55455}
}
\maketitle   
\begin{abstract}
We consider a (long-lived) sterile neutrino dark matter scenario in a five dimensional (5D) warped extra dimension model where the fields can live in the bulk, 
which is partly motivated from the absence of the absolutely stable particles in a simple Randall-Sundrum model. 
The dominant production of the sterile neutrino can come from the decay of the radion (the scalar field representing the brane separation) 
around the electroweak scale. 
 The suppressions of the 4D parameters due to the warp factor and the small wave function overlaps in the extra dimension help
 alleviate the exceeding fine-tunings typical for a sterile neutrino dark matter scenario in a 4D setup.

\end{abstract}
\setcounter{footnote}{0} 
\setcounter{page}{1}
\setcounter{section}{0} \setcounter{subsection}{0}
\setcounter{subsubsection}{0}

%\newpage
\section{Introduction}
We will discuss the cosmological aspects of the radion phenomenology
 in the Randall-Sundrum background with the fields in the bulk \cite{lisa1,gold,goldpot,riz,Csaki2,Csaki1,gro,tony3,csakic,csakia}. 
In particular we study the decay of the radion into the sterile right-handed neutrino which can account for all the dark matter in the Universe.
This is partly motivated from the absence of the discrete symmetry in a simple warped extra dimension model and consequently the lack
 of an absolutely stable particle which could be a promising dark matter candidate. 
This is in contrast to, for instance, a flat extra dimension model which possesses the discrete symmetry corresponding to the 
translational invariance in the extra dimension, called the Kaluza-Klein (KK) parity, leading to the absolutely stable KK dark matter \cite{tonydm,kk}
\footnote{Ref. \cite{aga} investigates the 
dark matter in a warped extra dimension model, where the discrete symmetry was introduced leading to the 
absolutely stable dark matter candidate.}.
We consider a simple addition of three right-handed neutrinos to the minimal 
Standard Model particle contents, where one of them is sterile (i.e. very weakly coupling to the other fields) 
with a lifetime longer than the age of the Universe (and the other two explain the atmospheric and solar neutrino data \cite{data} for the finite small neutrino masses).

The sterile neutrino can be either warm or cold dark matter in our scenarios.

The sterile neutrino warm dark matter in such a minimal extension of the Standard Model (sometimes referred to as $\nu$MSM \cite{asaka,numsm1,numsm4,tka} in a 4D setup) would have a potential interest from the astrophysical viewpoint because
it can ameliorate the shortcomings in the small (i.e. galactic) scale structures of the cold dark matter scenarios which the 
N-body numerical simulations apparently suffer from, such as the missing satellite problem (the cold dark matter models predict too many satellites 
(small dwarf galaxies about a thousandth the mass of the Milky Way) than observed) and the cusp/core problem (the 
cusped central density distributions rather than the observed smoother core in the dark matter haloes) \cite{prob1}.

One of the notable features in a warped extra dimension setup is the production mechanism of the sterile neutrinos which can come from the 
decay of the radion because the radion can couple to all the degrees of freedom. This would be of significant interest for a warm dark matter scenario where 
a simple Dodelson-Widrow ((non-resonant) active-sterile neutrino mixing \cite{scott1,asakamix}) mechanism alone cannot 
account for all the dark matter of the Universe due to the conflicts from the astrophysical observations such as the
 Lyman-$\alpha$ forest (giving the sterile neutrino dark matter mass lower limit of at least $\sim 10$ keV) 
and the X-ray data (giving the upper bound of at most $\sim 8$ keV) \cite{uros1,decaying,kev1,kevcold,boy1,john1,Dolgov1,mike1,silk1}.

The big fine-tunings, such as those typical for a sterile neutrino dark matter scenario in the 4D setups, are relaxed in the warped extra dimension 
models because the 4D parameters 
are exponentially suppressed naturally due to the warp factor and the small wave function overlaps in the extra dimension. We also mention that the desirable baryon asymmetry of the Universe in our simple warped extra dimension model can arise from the decay of the higher KK modes of the right-handed neutrinos \cite{lepto1}.

\section{Setup}
\label{setup}
We consider the scenario where the fields can live in the bulk of the 5D Randall-Sundrum spacetime whose $AdS_5$ metric in the conformally flat coordinate is \cite{lisa1}
\ba
ds^2=\left(\frac{R}{z}\right)^2(\eta_{\mu \nu}dx^{\mu}dx^{\nu}-dz^2)
\ea
where $z$ is in the interval $[R, R']$ with the UV (Planck) brane and the IR (TeV) brane respectively at $z=R,R'$. The radion corresponds to the scalar perturbation $F$ of the metric \cite{Csaki2,csakia,pert,krib2} \footnote{The general scalar perturbations about the background with the vacuum expectation value (vev) of Goldberger-Wise (GW) field $\phi_0$ \cite{gold,goldpot} is (bulk metric scalar fluctuations can be sourced by the perturbation of the GW field as well) 
\ba
\phi(x,z)=\phi_0(z)+\varphi(x,z ),~
ds^2=\left(\frac{R}{z}\right)^2 (e^{-2F(x,z)}\eta_{\mu \nu}dx^{\mu}dx^{\nu}-(1+G(x,z))^2 dz^2)
\ea
Plugging this parameterization ansatz into the Einstein equations reveals $G$ and $\varphi$ can be expressed in terms of $F$ (in particular $G=2F$ up to the linear order). Therefore we shall use only a single scalar mode $F$ to describe the scalar perturbations in our discussions.}
\ba
\label{metric}
  ds^2=
\left(\frac{R}{z}\right)^2(e^{-2F}\eta_{\mu \nu}dx^{\mu}dx^{\nu}-
(1+2F)^2 dz^2).
\ea
Assuming the form $F(x,z)=f(z)r(x)$, the canonically normalized radion $r(x)$ in 4D is obtained by
\ba
\label{canof}
F(x,z)=\left(\frac{z}{R'}\right)^2 \frac{r(x)}{\Lambda_r},~\Lambda_r\equiv \frac{\sqrt6}{R'}
\ea
The linear variation of the action 
\ba
\int d^5 x \sqrt{g} F(\mbox{Tr}T^{MN}-3 T^{55}g_{55})
\ea
can give us a coupling of the radion to the matter fields, where 
$g$ is the determinant of the 5D metric and we use 
the upper-case Latin (lower-case Greek) letters to denote the 5D (4D) indices.

Let us now consider a concrete example of a 5D bulk Dirac spinor consisting of two two-component spinors $\Psi^T=(\chi_{\alpha},\bar{\psi}^{\dot{\alpha}})$ whose standard 5D bulk action including the bulk Dirac mass term reads \cite{gro,tony3,csakic,csakia}
\ba
\label{action1}
\int d^5 x \left(\frac{R}{z}\right)^5 \left[\left(\frac{z}{R}\right)
\left(-i \bar{\chi}\bar{\sigma}^{\mu}\partial_{\mu}\chi-
i {\psi}\bar{\sigma}^{\mu}\partial_{\mu}\bar{\psi}
+\frac12(\psi \overleftrightarrow {\partial_5}\chi-\bar{\chi} \overleftrightarrow {\partial_5}\bar{\psi})\right) + m_D (\psi\chi+\bar{\chi}\bar{\psi})
\right]
\ea
where $(R/z)^5$ comes from the determinant and an additional factor $(z/R)$ comes from the 5D vielbein $e_a^M$. We shall use the standard parameterization of the bulk mass parameter $m_D=c/R$.  
Note that we did not perform the integration by parts for the 5D kinetic terms because, as we shall see in the following section, there arise the non-trivial boundary conditions due to the finite interval in the 5th dimension. By using the perturbations in the metric
\ba
\sqrt{g}=\left(\frac{R}{z}\right)^5 e^{-4F}(1+2F), ~
e_a^M=\mbox{diag}\frac{z}{R}(e^{F},e^{F},e^{F},e^{F},1/(1+2F))
\ea
the linear order couplings of the radion to a fermion for the above action can be obtained from \footnote{We here consider only the linear order coupling for simplicity. The higher order couplings such as $\bar{\chi}_n(x) \bar{\sigma}^{\nu} \chi_n(x)  \partial_{\nu} r(x) $ coming from the canonical normalization of the fermion field \cite{Csaki1} and the effects of the radion Higgs mixing through the curvature-Higgs couplings \cite{krib2,mix} would not affect the qualitative analysis even though they could potentially affect the quantitative discussions for some parameter range.} 
\ba
\label{radcoup}
\int d^5 x F\left(\frac{R}{z}\right)^4  
\left( i \bar{\chi}\bar{\sigma}^{\mu}\partial_{\mu}\chi+
i {\psi}{\sigma}^{\mu}\partial_{\mu}\bar{\psi} 
-2(\psi \overleftrightarrow {\partial_5}\chi-\bar{\chi} \overleftrightarrow {\partial_5}\bar{\psi})-\frac{2c}{z}(\psi\chi+\bar{\chi}\bar{\psi})
\right)
\ea

We are mainly interested in the radion couplings to the lightest 4D modes in the KK eigenstate decompositions 
\ba
\chi=\sum_n g_n(z)\chi_n(x),\psi=\sum_n f_n(z)\bar\psi_n(x)
\ea
where the 4D components satisfy the 4D equations of motion with the mass $m_n$ for each $n^{th}$ mode
\ba
-i \bar{\sigma}^{\mu}\partial_{\mu}\chi_n+m_n \bar\psi_n=0,
-i \sigma^{\mu}\partial_{\mu}\bar\psi_n+m_n \chi_n=0
\ea
In particular, for $n=0$, there are massless zero modes of form, with the standard Dirichlet boundary conditions $\psi|_{z=R,R'}=0$,
\ba
\label{waveno}
g_0=A \left(\frac{z}{R}\right)^{2-c},~f_0=0,~A=
\sqrt{\frac{1-2c}{R}}\frac{1}{\sqrt{\kappa^{2c-1}-1}},~\kappa\equiv \frac{R}{R'}
\ea
where the normalization constants are obtained by requiring the canonically normalized 4D kinetic terms. Eq. (\ref{waveno}) shows the fermion zero modes are localized around the Planck (TeV) brane for the bulk mass parameter $c>1/2$ $(c<1/2)$. The radion, on the other hand, is always localized to the TeV brane as can be seen by Eq (\ref{canof}). This means that, from the AdS/CFT correspondence \cite{nima1}, 
the radion is a composite state which shows up when the conformal symmetry is broken at the temperature around the TeV scale. 
We implicitly assume the confinement of the composites 
when we discuss the radion decay around the electroweak scale temperature.

\section{Sterile neutrino mass}
To obtain the abundance of the sterile neutrinos produced by the decay of the radion, we need to know the sterile neutrino mass 
in addition to the couping strength to the radion outlined in the last section.

We assume, for simplicity, three gauge-singlet right-handed neutrinos in addition to the minimal Standard Model particles, and 
consider the possibility for one of them, denoted as $N$, to be the sterile neutrino dark matter taking account 
of all the dark matter in the Universe. 
For it to be sterile, we assume the 4D Dirac Yukawa coupling of $N$ to the Higgs and left-handed neutrino is negligibly small (consequently its Dirac mass is also negligible compared with its Majorana mass) which could be justified by the suppression 
due to the warp factor and small wave function overlaps in the extra dimension (more quantitative discussions on how small it needs to be will be given in \S \ref{discu}). Because no symmetry prohibits the Majorana mass terms for the gauge singlet $N$, we can consider the lightest eigenmass of 
$N$ in existence of the brane localized Majorana mass term which dominates over the Dirac mass contributions.

We briefly review here, for the illustration purpose, the Majorana mass term confined on the Planck brane (we choose the basis where the Majorana mass is real) \footnote{The Majorana spinors cannot exist in 5D because of the lack of the real representation of $\gamma$ matrices \cite{gamma}, but we can add the gauge- and Lorentz-invariant bilinear terms for 5D Dirac isosinglet neutrino, $N^T C^{(5)-1}N$ ($C^{(5)}$ is a charge conjugation operator), which can be interpreted as the Majorana mass terms in 4D. Note 
each 5D Dirac spinor in general leads to two 4D Majorana states at each KK level in absence of the boundary conditions.} \cite{gro,tony3,csakic,sha,tonydirac}
\ba
\label{massterm}
 \int d^5 x \sqrt{g} \frac12 m_M (N_R N_R+h.c.), ~m_M=d_M \delta(z-R)
\ea
and the boundary conditions are accordingly modified to
\ba
\label{bdry}
N_L(z=R)=d_MN_R(z=R),~N_L(z=R')=0
\ea
where $d_M$ is a dimensionless constant and 
 \ba
N(x,z)= \left( \begin{array}{c}
N_L(x,z)\\
\bar{N}_R(x,z)
\end{array} \right), 
N_{L}=\sum_n f_{L}^{(n)} (z)N_L^{(n)}(x),
~\bar{N}_{R}=\sum_n f_{R}^{(n)} (z)\bar{N}_R^{(n)}(x)
\ea
We can however simplify our analysis by calculating the 4D mass eigenvalues/states via the diagonalization of the mass matrix in the basis obtained without the 
boundary Majorana mass $(N^{(0)}_{R},N^{(1)}_{R},N^{(1)}_{L},...)$. In this approximation, 
the symbols $f^{(n)}_{R,L},N^{(n)}_{R,L}$ are used for the states obtained without the Majorana mass terms 
rather than the mass eigenstates, and the lightest eigenstate dominantly consists of $N_R^{(0)}$ while $N_L^{(0)}$ is decoupled from the low-energy theory as the result of the boundary conditions. The approximate mass spectrum up to the $n^{th}$ KK level can be calculated 
by truncating at the $n^{th}$ level mass matrix when the mixing contributions from the higher modes are small \cite{sha, Dienes}. 
We are interested in the lightest 4D eigenstate (because of the boundary 
Majorana mass, the lightest mode now obtains a non-vanishing eigenmass $m_N$), and we omit the upper mode indices in the following unless stated otherwise. For instance, the lightest 4D eigenmass for the range of our interest $m_N\ll 1/z$ can be obtained, up to the leading order, by substituting 
the wave functions of Eq. (\ref{waveno}) into the brane localized Majorana mass term of Eq. (\ref{massterm}) as (note our use of $c$ is such that $c<1/2$ ($c>1/2$) for the localization to the TeV (Planck) brane) \footnote{We consider only $c<1/2$ because $m_N\approx 
d_M (2c-1)/R$ for $c>1/2$, and $|m_N z|\ll 1$ would require the fine-tunings in this case. Note we can obtain the same results as Eqs. (\ref{massc2}, \ref{tevmas}) by 
substituting the eigenstate solutions (in terms of the Bessel functions) of the 5D bulk equations of motion into 
Eq. (\ref{bdry}) and expansing the Bessel functions to the leading order for $|m_N z|\ll 1$.}
\ba
\label{massc2}
m_N\approx \frac{1}{R}{d_M}\left(\frac{R}{R'}\right)^{1-2c}(1-2c) \mbox{  for  }c<1/2~
\ea

The leading-order radion decay channel into the 
zero mode sterile neutrinos (approximated by ${N}_R^{(0)}(x)$) can be thus read off from 
Eq. (\ref{radcoup}) as \footnote{We here note that the radion couples only to the bulk terms and the 
brane-localized Majorana mass terms do not affect the radion couplings \cite{Csaki2}.
At first sight, one may naively expect that the Majorana mass terms of Eq. (\ref{massterm}) 
could lead to the radion coupling of
\ba 
\int d^4 x e^{-4F}\frac12 d_M (N_R N_R+h.c.)|_{z=R}
\ea
This radion coupling term however is exactly canceled by the surface term of the action (see Eqs. (\ref{action1}, \ref{radcoup})) ($[x]^{R'}_R\equiv 
x(R')-x(R)$)
\ba
\label{vari}
\delta S_{boundary}=\frac12\int d^4 x 
\left[\left(\frac{R}{z}\right)^4 e^{-4F}(\delta N_L N_R- \delta N_R N_L+h.c.)\right]^{R'}_R
\ea
because of the boundary conditions Eq. (\ref{bdry}).} 
\ba
\label{majocoup}
\int d^4x a_0 \frac{m_N}{\Lambda_r}r(x) \ {N}_R^{(0)}(x){N}_R^{(0)}(x)+...,~
a_0= 
\int dz \left(\frac{z}{R'}\right)^2\left(\frac{R}{z}\right)^4f_R^{(0)}(z) f_R^{(0)}(z)
\ea

The analogous arguments can be applied to the Majorana mass confined on the TeV brane, where the lightest eigenmass becomes 
\footnote{We consider only $c>1/2$ for the TeV brane confined Majorana mass case because $m_N\approx d_M (1-2c)/R'$ for $c<1/2$, and 
$|m_N z|\ll 1$ would require the fine-tunings.}
\ba
\label{tevmas}
m_N\approx \frac{1}{R'}{d_M}\left(\frac{R}{R'}\right)^{2c-1}(2c-1) \mbox{  for  }c>1/2
\ea

\section{Sterile neutrino abundance}

For an estimation of the abundance of the sterile neutrino dark matter, we integrate the Boltzmann equation 
($C_{col}$ is a collision term)
\ba
\frac{d n_N}{dt}+3Hn_N=C_{col}
\ea
which can be rewritten in terms of the so-called yield parameter $Y\equiv n/s$
\ba
\label{yield}
\frac{dY}{dT}=\frac{-C_{col}}{HTs}\left(1+\frac{T}{3g_s(T)}\frac{dg_s(T)}{dT}\right)
\ea
In the above, $n_N$ is the number density for the sterile neutrino, $s$ is the entropy density and $H$ is the Hubble constant given by, in terms of the 
radiation energy $\rho_R$,
\ba
H^2=\frac{8\pi G}{3}\rho_R,\rho_R=\frac{\pi^2}{30}g_* T^4, s=\frac{2\pi^2}{45}g_sT^3
\label{entro}
\ea
The total numbers of the effectively massless degrees of freedom are given by
\ba
g_{*}(T)=\sum_B g_B \left(\frac{T_B}{T}\right)^4
+\frac{7}{8}\sum_F g_F \left(\frac{T_F}{T}\right)^4,~
g_{s}(T)=\sum_B g_B \left(\frac{T_B}{T}\right)^3
+\frac{7}{8}\sum_F g_F \left(\frac{T_F}{T}\right)^3
\ea
where 
$g_{B(F)}$ is the number of helicity states for each boson (fermion) with their corresponding temperature $T_{B(F)}$. In deriving 
Eq.(\ref{yield}), we assumed the constant entropy and used Eq. (\ref{entro}). 

We note that the thermalization of the radion can be justified from the interactions involving the gauge couplings. For instance, its coupling to the gauge fields in 4D \cite{riz, Csaki2} can be obtained in the same manner as those outlined in \S \ref{setup}
\ba
\frac{Rg_4^2}{g_5^2}\frac{r(x)}{4\Lambda_r}F^{\mu \nu}(x)F_{\mu \nu}(x)\sim \frac{1}{4 \Lambda_r \log(R'/R)}
r F^{\mu \nu}F_{\mu \nu}
\ea
where $F$ is the field strength of the massless gauge field $A$ (photon or gluon) and $g_4, g_5$ are the 4D and 5D gauge couplings which are related via $g_5^2\sim g_4^2 R \log (R'/R)$. Note that the coupling constant has a suppression of the TeV scale which is the cut-off scale on the TeV brane. 
For example, the interaction rate of $r A \leftrightarrow ff$ around the TeV scale temperature is $\Gamma\sim  (e/\log(R'/R))^2\times TeV$ ($e$ is the electric charge of a fermion $f$) which indeed exceeds the Hubble expansion rate $H\sim (TeV/M_p)\times TeV$ (from the 4D Friedmann equation), so that we can use the thermal abundance for the radion.

Let us now parameterize the collision term as $C_{col}=n_r\gamma \lambda^2 m_r /8\pi$. Here $n_r,m_r$ denote the radion abundance and mass and $\lambda$ represents a dimensionless (4D) coupling constant between the radion and the sterile neutrino (in particular, this parameterization of the collision term corresponds to the decay channel $\lambda r NN$ to be discussed as our concrete examples in the next section). $\gamma$ is the mean time dilation factor in the decay and we use the Maxwell-Boltzmann statistics for the phase space distribution function $f(p)$
\ba
n_r(T)=\frac{g_*T^3}{2\pi^2}\left(\frac{m_r}{T}\right)^2K_2
\left(\frac{m_r}{T}\right), ~~
\gamma=\left \langle \frac{m_r}{E_r}\right \rangle =
\frac{\int d^3 p f(p)\frac{m_r}{E_r}}{\int d^3 p f(p)}=
\frac{K_1(m_r/T)}{K_2(m_r/T)}
\ea
where $K_{n}$ is the modified Bessel function of the second kind of order n (as expected, $n_r\propto g_* (m_rT)^{3/2}exp(-m_r/T),\gamma\sim 1$ for $T\ll m_r$ and $n_r\propto g_* T^3, \gamma\sim m/T$ for $T\gg m_r$) \cite{rocky1,gondolo1}. 

Then $Y$ can be obtained by integrating the Boltzmann equation
\ba
Y(x)&=&a\lambda^2 \frac{M_p}{m_r}y(x),~~y(x)=\int^x_{m_r/\mu_{TeV}} dx' K_1(x')x'^3, 
\ea
where $x=m_r/T$ and the numerical coefficient $a\sim 2.6 \times 10^{-4}$ with $g_*\sim 110$. The numerical integration of $K_1(x)x^3$ approaches quickly to the asymptotic value of $y(x\gg 1)\sim 4.7$ where we took the integration lower bound of $ m_r/\mu_{TeV}=0.1$ ($\mu_{TeV}$ is the scale of the conformal symmetry breaking) \footnote{There is an uncertainty for when the conformal symmetry breaking occurs, even though 
changing the lower bound to $m_r/\mu_{TeV}=0.01$ changes the asymptotic value less than $1\%$ and 
$m_r/\mu_{TeV}=1.0$ just slightly changes the asymptotic value to $y(x\gg 1)\sim 4.5$.}.

In the above integration, the time dependence of $g_s$ was assumed to be small. This assumption would be reasonable during the decay of the radion which dominantly occurs around $T\sim m_r$ (the electroweak scale). The most of the higher KK states which could possibly have a non-negligible effect on $dg_s(T)/dT$ are not energetically accessible around the conformal symmetry breaking energy scale when the composite particles appear. 
More detailed studies of the physics around the phase transition period will be presented in our future work 
\footnote{For instance, in our simple numerical estimations, we 
neglected the effects of the the thermal mass corrections whose significance would depend on the nature of the electroweak and conformal symmetry breaking phase transitions \cite{quiros}.}.
Matching $m_N Y$ to the 
current matter density value $\rho_{m}/s\sim 0.4 \times 10^{-9}$GeV \cite{pdg} gives an estimation of  
\ba
\label{abundance}
\lambda^2\sim 0.3\times 10^{-20}\left(\frac{1MeV}{m_N}\right)\left(\frac{m_r}{100 GeV}\right)
\ea

\section{Examples}
\label{eg}
Let us now discuss a few concrete examples for the illustration purpose.
For definiteness, we set $1/R$ to be the Planck scale and use the values for the 
radion mass and Majorana mass parameter as $m_{r}=300$ GeV and $d_M=1$. There then remain only two free parameters which are relevant for 
the dark mater abundance constraint Eq. (\ref{abundance}): $\Lambda_r(\equiv \sqrt{6}/R')$ and the bulk mass parameter 
$c_N$ for the bulk sterile (right-handed) neutrino. 

For the case of the Planck brane localized Majorana mass, we can, for instance, take $(c_N,\Lambda_r)\sim$ ($-0.28$, $10$ TeV) (corresponding to $1/R'\sim 4$ TeV) to satisfy 
Eq. (\ref{abundance}). This in turn gives $(m_N, \lambda)=(17$ keV, $7\times 10^{-10})$ for a warm dark matter candidate, which were obtained by integrating the 5D wave functions over the fifth dimension using Eqs. (\ref{massc2}, \ref{majocoup}) 
for the radion decay channel $r \rightarrow NN$ arising from the term 
$\lambda r(x) N(x) N(x)$. 
For the case of the TeV brane localized Majorana mass, the choice of $(c_N,\Lambda_r)\sim$ (0.63, 1 TeV) (corresponding to $1/R'\sim 0.4$ TeV), for instance, gives us $(m_N, \lambda)\sim (5$ MeV, $ 4\times 10^{-11})$ so that this case corresponds to a cold dark matter scenario. Note we could obtain these small 4D
parameters $m_N$ and $\lambda$ without fine-tuning our free parameters in our 5D model thanks to the 
exponential suppressions coming from the warp factors and 
small wave function overlaps in the extra dimension.

We also have the constraints from the Lyman-$\alpha$ forest analysis which gives a lower bound for the dark matter mass 
to ensure the enough small scale structures \cite{uros1}. 
The Lyman-$\alpha$ forest constrains the free-streaming length of the dark matter 
which in turn gives a constraint on the mass of the dark matter particles. The relation between the free-streaming length and the mass however depends on the production mechanism affecting the average momentum of the dark matter particles $\langle p_N \rangle$. For a simple estimation of 
the free-streaming length \cite{bond,earl}
\ba
\lambda_{FS}\sim 1Mpc\left(\frac{keV}{m_N}\right)\left(\frac{\langle p_N \rangle}{3.15 T}\right)|_{T\approx1keV}
\ea
For the sterile neutrinos produced from the active-sterile neutrino mixing mechanism, $\langle p_N \rangle \approx 2.8 T$ \cite{kevc}, while the average momentum for the relativistic fermions in equilibrium is $\langle p_N \rangle \approx  3.15 T$. The average momentum 
produced from the thermal radion decay has the average momentum $\langle p_N \rangle \approx 2.45 T$ \cite{tka}. 
Taking account of the additional redshifting of $\langle p_N \rangle$ by the 
decrease of the effective degrees of freedom $(g(T\sim$ TeV$)/g(T\ll$ MeV$) )^{-1/3}$, our scenario has 
$\langle p_N \rangle_{T\ll 1 MeV}  \approx 0.76 T$. 
We then find the Lyman-$\alpha$ constraints assuming the the active-sterile neutrino mixing mechanism for the sterile neutrino production
\cite{uros1}
\ba
m_N \gtrsim 10\sim 28 [keV]
\ea
correspond to \footnote{We here, for the number of degrees of freedom, used the Standard Model value $(g(T\sim $TeV$)/g(T\ll $MeV$) )^{-1/3}\sim 33^{-1/3}$.}, in our radion decay production scenario,
\ba
m_N \gtrsim  3 \sim 8 [keV]
\ea
The values realizable in our scenarios, such as those in our concrete examples, hence can be indeed consistent with the Lyman-$\alpha$ forest data, and they 
can also be consistent with another independent probe using the QSO 
gravitational lensing which gives the constraint $m_N \gtrsim 10$ keV \cite{miranda}.

The other astrophysical constraints, such as the 
observation of diffuse photon backgrounds \cite{decaying,kev1,boy1,Dolgov1,silk},
are related to the neutrino Yukawa coupling and consequently the mixing of the active and sterile neutrinos. 
Our scenarios can satisfy these astrophysical and cosmological constraints without affecting the dark matter abundance, 
simply because the dominant production of the sterile neutrinos in our scenarios can come from the radion decay and 
is unrelated to the active-sterile neutrino mixing \cite{scott1,asakamix,kevcold,john1,fuller} (also see Refs. \cite{tka,kusenkohiggs} 
for other production mechanisms). In fact, the desirable dark matter abundance is possible even with the vanishing neutrino Yukawa coupling for the sterile neutrino in our scenarios
.

\section{Discussion and conclusion}
\label{discu}
We showed that the sterile neutrino can be an 
interesting long-lived dark matter candidate in a simple warped extra 
dimension model which in general does not necessarily possess a discrete symmetry.

Although we assumed the (4D) neutrino Yukawa coupling $\lambda_{4D}$ for the 
sterile neutrino is negligibly small in our discussions, it may be still of possible interest to 
check how small it has to be. Let us for this purpose outline the constraints on the mixing angle $\sin \theta \sim \lambda_{4D}\langle H \rangle/m_N$ where $\langle H \rangle$ is the Higgs vev. 

There arises a constraint from the requirement of the negligible sterile neutrino production by the 
active-sterile neutrino mixings \cite{scott1,asakamix,kevcold}.
The fitting function for the sterile neutrino abundance produced from the (non-resonant) active-sterile neutrino mixing is
 \footnote{The mixing production occurs most efficiently at the temperature around $T_{max}\sim 133$ MeV$ (m_N/1 keV)^{1/3}$ \cite{scott1,bar}.}
\ba
\Omega_{N_{mix}}h^2 \sim 0.3 \left(\frac{\sin^2 2 \theta}{10^{-10}}\right) \left(\frac{m_N}{100keV}\right)^2
\ea
Another constraint comes from the long enough lifetime \cite{kevcold,Dolgov1,wdecay} by 
considering the decay into an active neutrino and two leptons $N\rightarrow {\nu}+l+\bar{l}$ via $Z$ boson exchange
\ba
\tau\sim 10^{15}\mbox{sec}\times \left(\frac{MeV}{m_N}\right)^5\left(\frac{10^{-10}}{\sin ^2 2 \theta}\right) 
\ea
which should be at least of 
order the age of the Universe $\sim 4 \times 10^{17}s$.\footnote{The decay channels of the sterile neutrino via the radion exchange do not affect our discussions 
because the radion interaction vertices are suppressed by $\Lambda_r\gg m_N$.}
There also exists the constraint from the diffuse photon backgrounds \cite{Dolgov1,mike1} due to the radiative decay into 
an active neutrino and a photon $N\rightarrow \nu \gamma$. The emitted monoenergetic photon has a narrow decay line whose width 
is determined by the Doppler broadening for a potential astrophysical probe. 
For instance, for $m_N\lesssim 20$ keV, the X-ray observatories such as Chandra and XMM-Newton \cite{decaying,kev1,boy1} can give a severer constraint than 
the lifetime constraint. 
The recent analysis of the XMM-Newton observation of the Andromeda galaxy \cite{decaying}, for example, indicates 
the mixing angle should be $\sin ^2 (2 \theta)\lesssim 10^{-11}$ 
for one of our examples $m_N\sim 17$ keV. 
Even though the more massive cold dark mass cases leading to the $\gamma$-ray backgrounds still 
have less precise data for detecting the dark matter decay lines, the recent analysis using the high-resolution spectrometer SPI set up on 
INTEGRAL satellite \cite{win,spi} can constrain the dark matter lifetime for $40$ keV$ \lesssim m_N \lesssim 14$ MeV and it constrains
the mixing angle to be less than $\sin ^2 (2 \theta)\lesssim 10^{-24}$ for $m_N\sim 5$ MeV. 
\footnote{There are other (less tight) constraints for the heavy sterile neutrinos coming from, 
for instance, Big Bang Nucleosynthesis, supernova and accelerator experiments \cite{bar,repdol,mevste,dolheavy,neww} and they
are satisfied for the small mixing angles consistent with the lifetime and photon background constraints.}

We hence see that our two examples $m_N=17$ keV for a Planck brane localized Majorana mass and $m_N=5$ MeV for a TeV brane localized Majorana mass in \S\ref{eg} require
 the 4D neutrino Yukawa couplings $\lambda_{4D}$ which can be obtained after integrating the wave functions over the fifth dimension 
to be respectively $\lambda_{4D}\lesssim 10^{-13}$ and $\lambda_{4D}\lesssim 10^{-17}$ (we took the Higgs vev $\langle H \rangle=246$ GeV). 
The free parameters in our model to determine $\lambda_{4D}$ are 5D neutrino Yukawa coupling $\lambda_{5D}$ and the bulk mass parameters $c_L$ for 
the left-handed neutrino (assuming the other parameter $c_N$ relevant for $\lambda_{4D}$ is already fixed from the dark matter abundance constraint as in \S\ref{eg}). 
It is easy to obtain such a small $\lambda_{4D}$ by localizing the left-handed neutrino toward the Planck brane (i.e. by letting 
$c_L>1/2$) due to the exponential suppressions of the wave function overlaps in the fifth dimension. We however found that $c_L$ is essentially 
constrained to be $1/2<c_L\lesssim 1$ in our simple scenarios to give the 4D (electron) Yukawa coupling of at least the order ${\cal O}(10^{-6})$. 
This is because the electron and the left-handed neutrino are in the same SU(2) doublet and they share the same $c_L$ (even though 
we can still freely vary the bulk mass parameter of the right-handed electron). Because of this constraint on $c_L$, to obtain 
$\lambda_{4D}\lesssim 10^{-13}$ ($\lambda_{4D}\lesssim 10^{-17}$) for a warm (cold) dark matter example, we would need our free parameter 
$\lambda_{5D}$ to be $\lambda_{5D}\lesssim 10^{-5}$ ($\lambda_{5D}\lesssim 10^{-6}$) where the relaxation of tunings compared with a 4D model 
by the order ${\cal O}(10^{-11}\sim 10^{-8})$ is a result of the exponential suppressions from the warp factor and the small wave function overlaps in the extra dimension. \footnote{ $\lambda_{5D}$, which 
is an arbitrary free parameter in our model, still however needs to be 
smaller than the naturally expected values of the order unity and its justification goes beyond the scope of the present paper 
(even though one might be tempted to introduce 
some symmetry for its justification (such as $U(1)$ as discussed for the $\nu$MSM \cite{numsm3}) for the vanishing $\lambda_{5D}$ or a small value coming from its breaking).}

The lightest mass eigenvalue for the light left-handed neutrino in the same generation as that of the 
sterile (dark matter) neutrino is much smaller than ${\cal O}(10^{-2})$ eV due to the negligible 
active-sterile neutrino mixing in our model, so that the atmospheric and solar neutrino data 
\cite{data} should be explained by the other two remaining right-handed neutrinos in our simple scenarios. 
We also mention that the desirable baryon asymmetry of the Universe can be induced in our model from the 
decay of the higher KK modes of the right-handed neutrinos \cite{lepto1}.

While most of the radion phenomenology and in particular 
the cosmology relevant for the radion has been studied for the fields confined on the brane, 
our study would hopefully open up an interesting possibility for the further investigations of both 
astrophysics and particle physics model building in a warped extra dimension with the fields in the bulk.

\subsection*{Acknowledgments}  
The author thanks T. Gherghetta, K. Olive, M. Peloso, Y. Qian, M. Shaposhnikov and M. Voloshin for the useful discussions. He especially thanks T. Gherghetta for the early stage of collaboration and the continuous encouragement.
This work was supported by DOE grant DE-FG02-94ER-40823.

%%%%%%%%%%%%%%%%%%%%%%%%%%%%%%%%%%%%%%%

\end{document}